\documentclass[conference]{IEEEtran}
\IEEEoverridecommandlockouts
\usepackage{cite}
\usepackage{amsmath,amssymb,amsfonts}
\usepackage{algorithmic}
\usepackage{graphicx}
\usepackage{textcomp}
\usepackage{xcolor}
\usepackage{multirow}
\usepackage[symbol]{footmisc}
\usepackage[colorinlistoftodos]{todonotes}

\def\BibTeX{{\rm B\kern-.05em{\sc i\kern-.025em b}\kern-.08em
    T\kern-.1667em\lower.7ex\hbox{E}\kern-.125emX}}
\begin{document}

\newcommand\copyrighttext{%
  \footnotesize \textcopyright 2024 IEEE. Personal use of this material is permitted. Permission from IEEE must be obtained for all other uses, in any current or future media, including reprinting/republishing this material for advertising or promotional purposes, creating new collective works, for resale or redistribution to servers or lists, or reuse of any copyrighted component of this work in other works. 
  \linebreak Preprint submitted to the 10th IEEE International Smart Cities Conference 2024 (ISC2 2024).
  }
\newcommand\copyrightnotice{%
\begin{tikzpicture}[remember picture,overlay]
\node[anchor=south,yshift=10pt] at (current page.south) {\fbox{\parbox{\dimexpr\textwidth-\fboxsep-\fboxrule\relax}{\copyrighttext}}};
\end{tikzpicture}%
}

\title{When Circular Economy Meets the Smart City Ecosystem: Defining the Smart and Circular City}
\makeatletter
\newcommand{\linebreakand}{%
  \end{@IEEEauthorhalign}
  \hfill\mbox{}\par
  \mbox{}\hfill\begin{@IEEEauthorhalign}
}
\makeatother

\author{
\IEEEauthorblockN{Georgios Mylonas\thanks{This work was partially supported by the RRREMAKER project, funded by the European Union Horizon 2020 Research and Innovation program under the Marie Sklodowska Curie - RISE grant agreement n$^o$ 101008060, and by the European Union under the Italian National Recovery and Resilience Plan (NRRP) of NextGenerationEU, partnership on ``Telecommunications of the Future'' (PE00000001 - program RESTART - CUP E83C22004640001) in the WITS project and the project SERICS (PE00000014 - CUP D33C22001300002).}}
\IEEEauthorblockA{\textit{Industrial Systems Institute} \\
\textit{Athena Research Center}\\
Patras, Greece\\
ORCID: 0000-0003-2128-720X\\
}
\and
\IEEEauthorblockN{Athanasios Kalogeras}
\IEEEauthorblockA{\textit{Industrial Systems Institute} \\
\textit{Athena Research Center,}\\
Patras, Greece\\
kalogeras@isi.gr}
\and
\IEEEauthorblockN{Sobah Abbas Petersen}
\IEEEauthorblockA{\textit{Dpt. of Computer Science} \\
\textit{Norwegian University of Science and Technology}\\
Trondheim, Norway\\
sap@ntnu.no}
\linebreakand
\IEEEauthorblockN{Luis Muñoz}
\IEEEauthorblockA{\textit{Dpt. of Communications Engineering} \\
\textit{Universidad de Cantabria}\\
Santander, Spain\\
luis@tlmat.ucan.es}
\and
\IEEEauthorblockN{Ioannis Chatzigiannakis}
\IEEEauthorblockA{\textit{Dpt. of Computer, Control and Management Engineering} \\
\textit{Sapienza University of Rome}\\
Rome, Italy\\
ichatz@diag.uniroma1.it}
}


\maketitle

\copyrightnotice

 
\begin{abstract}
Smart cities have been a very active research area in the past 20 years, while  continuously adapting to new technological advancements and keeping up with the times regarding sustainability and climate change. In this context, there have been numerous proposals to expand the scope of smart cities, focusing on resilience and sustainability, among other aspects, resulting in terms like smart sustainable cities. At the same time, there is an ongoing discussion regarding the degree in which smart cities put people at their centre. In this work, we argue toward expanding the current smart city definition by integrating the circular economy as one of its central pillars and adopting the term smart (and) circular city. We discuss the ways a smart and circular city encompasses both sustainability and smartness in an integral manner, while also being well-positioned to foster novel business activity and models and helping to place citizens at the heart of the smart city. In this sense, we also argue that previous research in smart cities and technologies, such as those related to Industry 4.0, can serve as a cornerstone to implement circular economy activities within cities, at a scale that exceeds current activities that are based on more conventional approaches. We also outline current open challenges in this domain and research questions that still need to be addressed.
\end{abstract}


\section{Introduction}

We are currently at a global crossroads, on the one hand experiencing first-hand the effects of climate change and grasping the necessity of swift action to counter it, while on the other hand, gradually realizing the limits of current economic models in terms of sustainability and the need to change our production and consumption patterns at a global scale. At the same time, as stated by the United Nations\cite{un-cities-report} ``cities are here to stay, and the future of humanity is undoubtedly urban''; the current urbanization rate of 51\% (2021) will continue to grow and is expected to reach 65\% by 2050. Moreover, cities generate more than 80\% of global GDP, while they also produce 70\% of global greenhouse emissions and 50\% of global waste. This means that cities have an outsized effect globally on our present and our future, which to a great extent  justifies the great amount of research activity invested in smart cities during the past two decades.

In this setting, there has been a rapid increase of interest towards the concept of the circular economy, and whether and how it can replace the linear economy model we currently use. The concept of the Circular Economy (CE) is still being actively shaped and defined\cite{circular-economy-definitions}, since it entails a myriad of forms and ways to implement it around the world. Moreover, given the above, the processes taking place inside cities are a crucial part of the equation towards moving to a circular society globally. Currently, circular economy processes can be and are, in the majority of cases, implemented without the use of technology; we cannot stress enough how important it is to continue to use such existing circular economy chains, since i) they have been formed via years of answering existing problems in a practical manner, and ii) they are heavily based on the involvement of citizens to, e.g., collect, sort and repurpose materials and artefacts, resulting in creating or revealing new communities and value chains. As technologists and engineers, we believe that there are many things to learn from such paradigms, and that it is imperative to study them more closely as ``lessons'' on how to put people at the centre of smart cities, an objective often stated by the smart city community that is not achieved as often as it should be. However, in our opinion, we should also start looking more actively as a community into the integration of existing and novel technologies within the domain of the circular economy in order to enable new applications and opportunities, and more specifically when discussing the implementation of circularity within urban areas. Technologies could allow circular economy networks to scale better within the boundaries of cities, as well as to enable the discovery of hidden value chains that are otherwise difficult to spot. 

Given the above, \textit{circular cities} have been gradually entering the discussion more systematically, opening a discussion on how to integrate the circular economy within cities. Smart cities have been on the research agenda for decades and they are heavily based on digitalization, while the realization of the circular economy also depends on digitalization to a certain extent. There is a wealth of results related to smart cities and aspects such as data-driven decisions, predictive data analytics and real-time monitoring loops, as well as implementations of systems supporting waste management and smart manufacturing. These are all aspects and results that could be transferred and applied to the domain of circular economy within cities. Furthermore, in terms of governance, cities can be more flexible compared to central governments; in this sense, it is possible that \textit{a new vision for smart cities, i.e. the smart and circular cities}, can provide an effective means to accelerate the transition towards a circular economy model.

In other words, it appears as if smart cities and the circular economy could be a good match. We argue here that while ``circularity'' is an aspect in smart cities that has not received much attention from the technical community until now, it could fit very well with the current smart cities research activity. We summarize the current landscape with respect to smart cities, the circular economy and circular cities, and then posit our definition of the smart and circular city, followed by a discussion on several dimensions related to the integration of existing smart city research over the circular economy domain. We also provide a discussion on the challenges related to this vision and a discussion on the future of smart cities.

\subsection{Related Work}

Smart cities have attracted a wealth of attention from the research community in the past decades. As a research field that keeps evolving with the adoption of new technologies and the integration of new aspects, there have been several attempts to redefine the vision for smart cities and what we should be aiming for and how to implement such a vision. A recent example of a position paper attempting to redefine what we think about smart cities as a term was \cite{smart-city-21st-century}, where the authors discuss the evolution of the smart city concept, noting that the ``smart city'' term replaced the previous ``digital city'' in the late 2000s, while also briefly discussing characteristics usually attributed to smart cities. \cite{dejong2015} is another work discussing the multitude of definitions and approaches that have surfaced in the 2000s and 2010s with respect to the evolution of cities and urbanization, focusing on the twelve most frequent city categories in the bibliography up to 2013. The authors argued that there were 6 categories that are relatively distinct and ``supported by a specific body of theories''.

In addition, there have been some initial attempts to define the interplay between the domains of smart cities and the circular economy, such as in \cite{smart-circular-cities-mdpi}, where the author focused on aspects related to governance and the digital/green transition, using examples from the region of Tampere, Finland. It is also mentioned that the urban metabolism concept is ``supplemented by the view of digital technologies, thus creating a link to a smart city discourse''. Another recent work\cite{moving-towards-the-circular-economy} analyzed the concept of the circular city and discuss the tools towards implementing such a city. An in-depth discussion on related indicators that could be used as an evaluation framework for the effectiveness of a circular city is also included, with the authors surveying the current state-of-the-art and proposing additional indicators specifically for circular cities.

The extensive set of definitions given to circular economy in recent years is surveyed in \cite{circular-economy-definitions}, where a total of 221 related definitions is discussed and whether recent research has resulted in a shared understanding of what the circular economy term means. The authors argue that in recent years there is evidence pointing to a consolidation of some core principles, together with new research branches making their appearance, while also noting that much of the recent circular economy-related research lacks applicability.

The connection between the circular economy and Industry 4.0 is discussed in \cite{connecting-ce-and-industry4}. The authors identify a number of enablers linking these 2 domains, as well as related barriers that need to be dealt with when applying Industry 4.0 methods in a circular economy context. Such enablers and barriers can be considered for modeling and optimization purposes in this context. In a similar fashion, \cite{synergy-between-ce-and-i4-0} provides a survey on the synergies between these two domains, discussing the scarcity of work utilizing Industry 4.0 for implementing circular economy aspects, and especially regarding the assessment of the impact of the transition to a circular economy. Furthermore, the use of digital twins and smart manufacturing within the context of smart cities is discussed in \cite{city-digital-twins}, where it is argued that the application of digital twins for circular economy and sustainability in a smart city context currently appears to be quite limited. Finally, challenges related to the implementation of looping actions (``reuse, recycling and recovery of resources'') are discussed in \cite{circular-cities-challenges-to-implementing-looping-actions}, where 58 challenges are identified ranging from cultural, regulatory, institutional and educational, to technical and political.

In this work, we argue and go beyond previous related work by focusing specifically on the interaction between existing smart city research and how it can facilitate the transition to a circular economy, while also providing a definition for the smart and circular city term considering these dimensions.

\section{Towards Defining the Future of Cities}

Definitions for the circular economy are currently abundant; \cite{circular-economy-definitions} surveyed 221 definitions of the circular economy, noting that this multitude of definitions can at times create confusion. A relevant compact definition has been provided by the European Parliament\cite{circular-economy-europarl}, in which the ``circular economy (CE) is a model of production and consumption, which involves sharing, leasing, reusing, repairing, refurbishing and recycling existing materials and products as long as possible. In this way, the life cycle of products is extended''. 

While definitions for circular cities are not as abundant, several recent works attempted to bring clarity to the term. A circular city is defined in \cite{prendeville-2018-six-cities} as ``a city that practices CE principles to close resource loops, in partnership with its stakeholders (citizens, community, business and knowledge stakeholders), to realize its vision of a future-proof city''. In addition, the European Investment Bank argues in \cite{eib-ce} that ``(a circular city) is not the sum of its circular activities. It must also fully realize and exploit its potential as a cradle for circular development and use its governance tools and levers as catalysts for circular change''.

The Ellen MacArthur Foundation has provided a more detailed definition\cite{macarthur-foundation-circular-cities}, where a ``circular city has embedded the principles of the circular economy across the entire urban area. Everything is operating within an interconnected network of systems that are designed to eliminate waste and pollution, circulate products and materials, and regenerate nature. There’s collaboration between citizens, government, research facilities, and businesses. And the entire city is powered by renewable energy. In practice this means products, services, infrastructure, buildings, and vehicles are designed to be durable, adaptable, modular, easy to maintain and repurpose, and locally sourced. Everything can be composted, reused, or recycled. Nature is flourishing, abundant, and used as design inspiration. The result is a thriving local economy that provides a vibrant, livable and resilient way of life.''

Another definition is provided within the European Circular Cities Declaration\cite{circular-cities-declaration}: ``A circular city is one that promotes the transition from a linear to a circular economy in an integrated way across all its functions in collaboration with citizens, businesses and the research community. This means in practice fostering business models and economic behaviour which decouple resource use from economic activity by maintaining the value and utility of products, components, materials and nutrients for as long as possible in order to close material loops and minimise harmful resource use and waste generation. Through this transition, cities seek to improve human well-being, reduce emissions, protect and enhance biodiversity, and promote social justice, in line with the Sustainable Development Goals.''

\section{So, What is A Smart and Circular City? A Definition}

In most cases, previous uses of the ``smart circular city'' term did not provide clear, concise definitions of what the term corresponds to, in contrast with the plethora of definitions offered for the circular economy and circular cities. In one of the first mentions of the term ``Smart and Circular City'', the role of co-creation, citizen participation and marketplaces is brought forward in \cite{ichatz-luis-circular}, as examples of existing smart city results that could aid in the integration of circular aspects, without however giving a concrete definition. In \cite{smart-circular-cities-mdpi}, although there is a lengthy discussion on several aspects tied to the implementation of circularity within a smart city, it is only briefly mentioned that the  ``smart circular city (SCC) builds on the idea of a circular economy in the urban context, with a focus on digitally enhanced smartness''. In our opinion, although the term has started to surface in recent years as combination of the smart city and the circular economy, there is currently a lack of clarity as regards what the term could mean and what is the vision for a smart circular city.

This comes in contrast with the importance and role of the circular economy for the evolution of the smart city vision. Having the above in mind and trying to combine the current views on smart cities and the circular economy, we propose the following definition for smart and circular cities:

\textit{A smart and circular city leverages Information and Communication Technologies (ICT) to enhance its livability, sustainability and resilience, while also meaningfully integrating circular economy principles into its digital ecosystem to minimize waste and maximize productivity and resource reuse. This convergence of technologies with the circular economy creates an urban system-of-systems where data-driven decision-making, efficient energy use, intelligent transportation systems and interconnections between the digital and the physical world are employed to optimize city operations and services, as well as citizens' satisfaction, health and well-being. In addition, circular economy practices such as remanufacturing, digitalization of material/value flows, urban mining, along with community engagement are integrated to the city's technological backbone to move towards a sustainable, resilient, healthy and circular urban ecosystem. The convergence between the technological enablers and the circularity paradigm is also taken into consideration at a policy/regulatory level, creating a durable and transferable model.}

In this sense, such a city can reach a greater level of sustainability by integrating circular economy aspects, while also helping accelerate the transition towards circular economy, to a certain extent, by utilizing smart city research. In Fig.~\ref{fig:diagram}, we depict a mapping of several smart city research dimensions over the types of flows within the circular economy. Using the multi-flow metabolism concept presented in \cite{blomsma2023making}, which contains 4 different types of flows (material, energy, information and value), \textit{we also position citizens at the core of both the smart city and the circular economy}, and place six dimensions of existing smart city research across these types of flows, seen as spanning across this domain. An additional circular dimension is added at the exterior of the circle, with Planning, Implementation and Assessment describing the dynamicity that can be contributed by smart cities to the circular economy, enabling dynamic feedback loops. Our vision on how these dimensions fit within the circular economy context follows.

\begin{figure*}[t]
    \centering
    \includegraphics[width=0.52\textwidth]{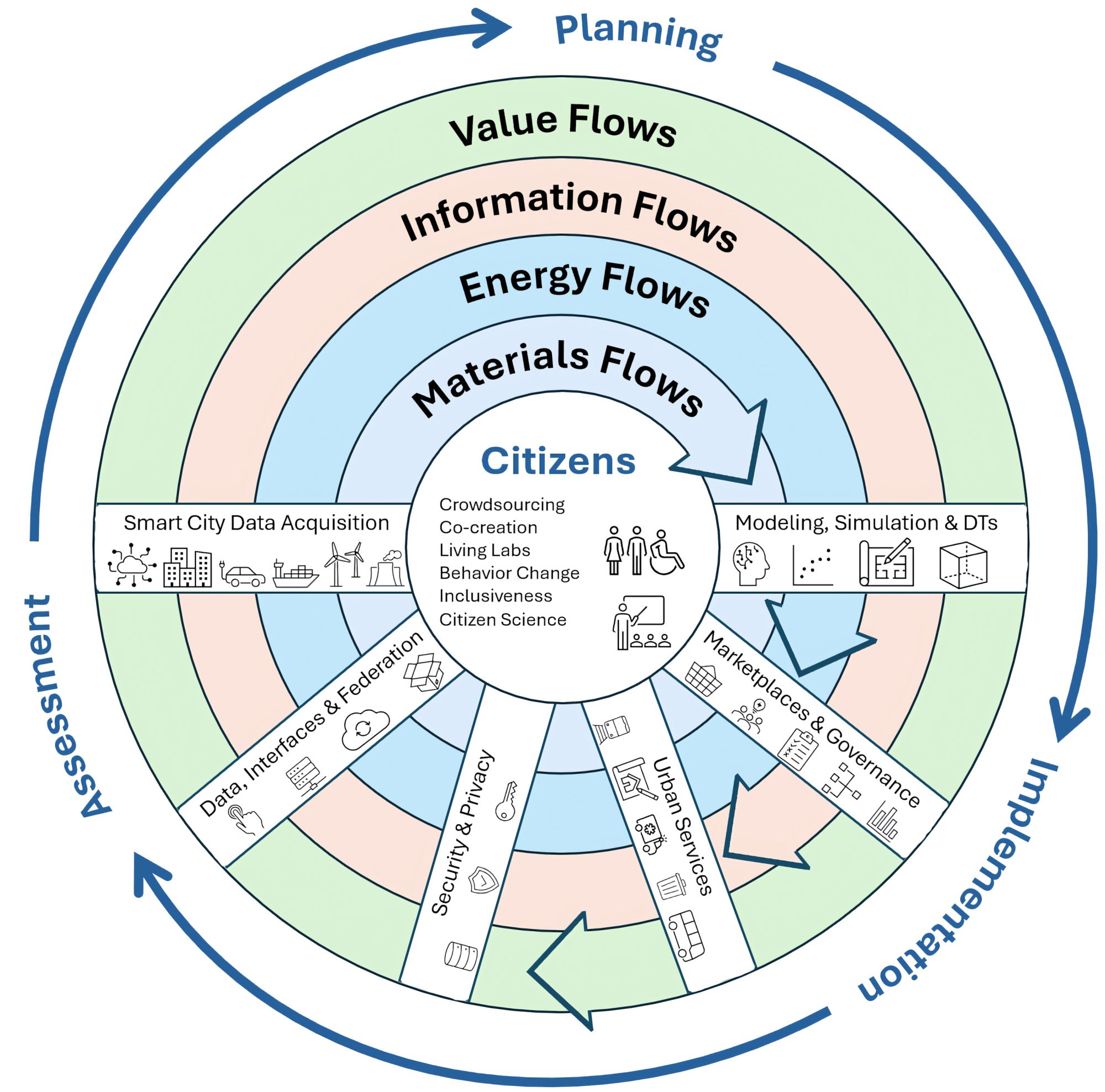}
    \caption{Integration of existing smart city research over the circular economy domain. Citizens are at the core of the smart city and the circular economy, with 6 smart city dimensions being applied over 4 types of circular economy flows, enabling a dynamic loop for continuous planning, implementation and assessment within a smart and circular city.}
    \label{fig:diagram}
\end{figure*}

\noindent\textbf{Citizens:} a great deal of smart city research has focused on citizens and  engagement in related activities, such as on crowdsourcing, co-creation and living labs. Such activity could be utilized in a circular economy context to boost their active involvement, while citizens can also act as low-cost sensors complementing sensing infrastructure. Furthermore, there is a body of work on behavior change and sustainability, which is a fundamental aspect of the transition to a circular economy, together with work on inclusiveness, citizen science, education, all aspects that enable the ``just transition'', an aspect often overlooked in both smart city and circular economy research.

\noindent\textbf{Smart City data acquisition:} smart cities are based on the bridges between the real and the digital world provided by sensors and IoT. There is a plethora of sensing infrastructure in smart cities from smart buildings, smart mobility, electric car charging points, smart ports, renewable energy production facilities, and pollution monitoring stations, to name a few examples. Such infrastructure can be directly utilized to monitor e.g., material and energy flows within a smart circular city, allowing new capabilities and feedback loops for monitoring. 

\noindent\textbf{Data, Interfaces and Federation:} data interconnections and APIs between applications and systems have long been an area for research in smart cities, with a number of solutions also facilitating federation between smart city platforms. Together with open data, these can enable a data economy. Moreover, human-computer interaction research concerning smart city analytics can  provide a good basis for user interfaces. Such results can \textit{empower data exchange for circular economy flows locally and between cities}, augmenting the scope of existing activities, while  providing more user-friendly interfaces.

\noindent\textbf{Security and Privacy:} as soon as digitalization and uncovering the value of smart city data enter the discussion, the importance of \textit{having security and privacy aspects considered and integrated within smart and circular city systems} cannot be understated. This should be a horizontal dimension across platforms; in smart cities, such aspects have already been tested and used, and could thus provide the underpinnings for secure implementations of circular economy activities. Distributed Ledger Technologies (DLT) have been used as tools to notarize data usage, enabling in this way \textit{data integrity and traceability}. There is also the dimension of security and privacy when training AI/ML models, together with the right to be forgotten. Methods like Federated Learning, already applied in smart cities, could augment the scope of circular economy processes, without compromising security and privacy.

\noindent\textbf{Urban/Smart City Services:} a range of services including public transport, waste management, health provision and urban planning, have been integrated within smart city platforms. In many cases, circular economy flows coincide with the aspects monitored in such smart city systems: from materials in waste management and energy flows in public transport, to construction materials of buildings being considered for urban planning, already implemented smart city services have the potential to support circular economy initiatives decisively.

\noindent\textbf{Marketplaces and Governance:} on the one hand, existing smart city marketplaces can support sustainable business models and potentially help to expand the current degree of involvement of the business sector. On the other hand, existing smart city governance mechanisms for policies, regulations and strategic agendas can help face some of the greatest challenges towards making the transition to a circular economy.

\noindent\textbf{Modeling, Simulation and Digital Twins:} this aspect can be viewed as the opposite end of smart city data acquisition, i.e., to be able to model entire cities and flows within them in a systematic manner. This is a capability with the potential to transform the way circular economy is implemented and monitored, facilitating feedback loops and affecting smart city policies, regulations and agendas. In particular, AI/ML-based methods can help translate the scale achieved through the smart city data ecosystem into truly dynamic models of cities.

\section{Open Challenges}

In terms of open challenges for smart and circular cities, there are several important aspects that need to be addressed. First of all, as is evident from this work, there is a wealth of definitions and views on what circular cities are and should be, while there is a lack of clarity as to how such visions take into account the existing smart city landscape. As a community, \textit{we need to define a vision and strategy that is rooted in both existing smart city results and circular economy}, but also utilize the potential synergies between these two domains, in order to facilitate the transition to a more sustainable future.

In this context, an additional open challenge is to define the \textit{role of smart city platforms for measuring circular economy flows within cities}, since smart cities offer great opportunities in this sense, while also defining specific indicators. Several attempts have produced lists of indicators to monitor and assess the performance of cities, such as the KPIs for Smart Sustainable Cities defined by United for Smart Sustainable Cities (U4SSC) UN initiative \cite{u4ssc}, or the indicators discussed in \cite{moving-towards-the-circular-economy}. However, such efforts should also take into account the smart and circular city point-of-view, in order to produce a framework that will help to assess the integration of circular economy aspects specifically in smart cities.

In addition, instead of working in silos, another open challenge is to \textit{bring different players in the smart city and circular economy domains together via digital interfaces and standardization}. Digitalization through the use of marketplaces and interfaces between existing platforms should help overcome such obstacles. The challenge of standardization is directly related to this dimension, while also being open for smart cities as well. This will also enable the creation of feedback loops to facilitate assessment with relevant KPIs. An additional dimension in this case would be to pair in particular the corresponding KPIs with AI/ML methods, aiming at providing a good balance, while also considering not just a single city but the \textit{federation of correlated cities}, e.g., due to geographical, social or other parameters. These cities could share data through a common platform in order to maximize synergies and achieve greater scale in sustainability-related results.

Furthermore, there is the challenge of \textit{successfully integrating industrial symbiosis and defining the role of Industry 4.0 in this domain}. In other words, whether existing approaches/know-how from Industry 4.0 research can be utilized as guides to drive adoption of circularity inside cities~\cite{blomsma2023making}. The flow of resources is another key aspect of circularity in cities and in industry. Key resources in smart cities include technological solutions and enablers, data, information, energy, knowledge and materials. For example, building materials could be reused in building and construction, data gathered in one project could be used across many situations and the energy produced in one process is used for enabling others, e.g., as thermal energy. Most importantly, the experiences and lessons learned from one situation could be a valuable resource in others, especially the knowledge and wisdom of the people that are involved.

As discussed in \cite{connecting-ce-and-industry4} from the point of view of enablers and barriers, the main identified Industry 4.0 enablers to enhance circularity comprise AI/ML, grouping together such technologies as visual computing and blockchain, and Circular Economy comprising Cyber-physical production systems, IoT, waste and energy recovery. Such enablers contribute to better circular strategies allowing product lifecycle extension, industrial waste optimization, supply chain sustainability enhancement, and overall performance monitoring. On the other hand, the main barriers for an I4.0-enabled Circular Economy are Interface design, grouping together design, investment cost, compatibility, interfacing and networking, and Automated Synergy model, combining CPS standardization, specification and modeling, automation system virtualization, process automation and digitalization, and collaborative models. Such barriers hinder the integration, interoperability and automation needed in the context of an ecosystem to ensure flexibility and agility. Increasing the level of integration and automation can help understand material, energy, information and value flows, like in the case of a multi-flow metabolism common framework approach \cite{blomsma2023making}. In addition, and perhaps even as importantly, \textit{such processes can aid in uncovering hidden value chains, that are otherwise very difficult to detect}. 

Additionally, \textit{Sustainable Business Models} have been driven by the desire to achieve sustainable solutions that respect the supply of natural resources and preserve our cultural and social values while making money. They can raise awareness about the resources that are used, produced and/or wasted in situations. Sustainable business models are discussed in \cite{triple-layered-business-model-canvas}. This is an essential part of a smart and circular city and the creation of a sustainable business model could facilitate circularity through the identification of the resources that could flow from one situation to another or within specific situations. Furthermore, there are additional challenges in relation to \textit{regulatory frameworks that can drive the adoption of circular economy processes within smart cities}, e.g., by imposing restrictions to the use/reuse of materials in specific sectors, such as the construction sector.

In terms of societal and citizen engagement aspects, people should be at the center of smart and circular cities. Aspects such as \textit{increasing the sustainability and circularity awareness of citizens}, together with \textit{supporting their engagement and participation in co-creation processes}, as well as \textit{having inclusivity embedded} within such activities, are still open challenges for the vast majority of cities.

There is also the challenge of harnessing the potential of the new capabilities offered by generative AI and machine learning within the circular economy context. Machine learning-based methods can facilitate use of the swaths of data inside smart cities, e.g., by identifying possibilities in reusing items or materials, at scale, within remanufacturing or upcycling scenarios.

Finally, the adoption of circular practices within smart cities, in general, is still not widespread. In terms of actual cities being categorized as circular, \cite{moving-towards-the-circular-economy} provided a list of circular cities in 2019. This list has grown considerably, as evidenced by activities such as the European Union's Circular Cities and Regions Initiative \cite{circular-cities-regions}. In addition, the EU Cities mission aims to ``deliver 100 climate-neutral and smart cities by 2030''\cite{eu-mission}. However encouraging these initiatives may be with respect to the transition to a circular economy, one could argue at this stage that in the majority of cases they are focused mostly on declarations of targets to be met in the near future and the related self-imposed deadlines, which are expressed in a uniform manner and do not necessarily have the same context.

\section{Discussion and Conclusions}

The conversation about integrating circular economy processes within the urban fabric has gained traction in recent years. Several approaches and theoretical frameworks has been proposed in the context of shaping the future of cities and our society. To a certain extent, the discussion about circular economy has been done in parallel with the discussion on smart and sustainable cities, with the interaction between these two domains/communities being relatively low, or at least not corresponding to the potential of smart and circular cities. However, we should also keep in mind that there are a lot of existing circular economy practices that work, all around the world. In our opinion, as a technical community we should not aim for replacing such existing practices that are successful, but instead we should strive for learning from their successes and transfer their strong points over to smart and circular cities.

There are several questions that need to be answered in the context of smart and circular cities. One such question is whether the circular economy concept is well-suited to technology-based solutions, or, in other words, if we want to view cities as systems-of-systems or a giant digital twin, maybe we should be focusing more on circularity as a technical community. On the one hand, there are multiple dimensions to consider what is circular within a smart city: data, technologies, solutions, knowledge, among others. On the other hand, viewing this from the opposite side of the argument, the question would be what do smart city technologies bring onto the table with respect to circularity and to what extent can a city be circular without also being smart? 

Furthermore, on a business level, another question to be answered is whether smart cities should start considering the potential of the circular economy as a means to create more business opportunities. In addition, the potential of using Industry 4.0 processes to identify hidden value chains is also a strong argument in favor of moving towards smart and circular cities. Finally, another question is what is the current level of awareness of citizens regarding sustainability and circularity, and how can smart cities help in increasing such awareness, or help to change the citizens' behavior towards more sustainable and circular consumption patterns.

\bibliographystyle{IEEEtran}
\bibliography{bibliography}

\end{document}